\documentclass[fleqn,10pt]{wlpeerj}

\usepackage{amssymb}
\usepackage{hyperref}
\usepackage{natbib}
\usepackage{float}

\title{Schelling segregation dynamics in densely-connected social network graphs}

\author[1]{Sage Anastasi}
\author[2]{Giulio Valentino Dalla Riva}
\affil[1]{University of Canterbury, Private Bag 4800, Christchurch 8140}
\affil[2]{University of Canterbury, Private Bag 4800, Christchurch 8140}
\corrauthor[1]{Sage Anastasi}{sage.anastasi@pg.canterbury.ac.nz}

\begin{abstract}
Schelling segregation is a well-established model used to investigate the dynamics of segregation in agent-based models. Since we consider segregation to be key for the development of political polarisation, we are interested in what insights it could give for this problem. We tested basic questions of segregation on an agent-based social network model where agents' connections were not restricted by their spatial position, and made the network graph much denser than previous tests of Schelling segregation in social networks.

We found that a dense social network does not become as strongly segregated as a sparse network, and that agents' numbers of same-group neighbours do not greatly exceed their desired numbers (i.e. they do not end up more segregated than they desire to be). Furthermore, we found that the network was very difficult to polarise when one group was somewhat smaller than the other, and that the network became unstable when one group was extremely small; both phenomena may help explain the complexity of real-world polarisation dynamics, such as unique risks faced by very small groups in a society. Finally, we tested Fossett’s \cite{fossett_preferences_2006} “paradox of weak minority preferences”, a well-established result in grid- and map-based models which shows that an increase in the minority group’s desire for same-group neighbours can create more segregation than a similar increase for the majority group. In a densely connected social network, we find that the evidence for this effect is mixed.
\end{abstract}

\begin{document}

\flushbottom
\maketitle
\thispagestyle{empty}

\section*{Introduction}

We are interested in developing methods of measuring polarisation that do not focus on political parties. This is because we believe that the polarising division may occur around issues that do not map directly to political party stances, and may not be captured well by voter data. For example, voter data in the USA does not capture the views of Black people before 1964, and analyses which use it are therefore missing an important part of the society (\cite{hetherington_review_2009}). Furthermore, methods which focus on political parties require sorting the parties into a left-right ideological divide, and this is difficult in the New Zealand case. The populist New Zealand First party frequently receives enough of the vote to be placed in a "kingmaker" position in coalition negotiations, and has entered coalitions with both left- and right-wing parties during its history; in 2017 it formed a coalition with the left-wing Labour party, and in 2023 it formed a coalition with the right-wing National and Act parties. This means that changes in New Zealand's political environment cannot be captured just by a left-right divide, and the political field cannot be simplified by excluding New Zealand First due to their frequent importance in coalition formation. Furthermore, there are several unique factors in the New Zealand political landscape that affect the country's susceptibility to populist politics (\cite{vowles_exception_2020}), which means that we may benefit from new methods to detect populism.

Our previous work developed an indicator of polarisation based on the embedded dimensionality $\hat{d}$ of a network of connections between the users, i.e. the number of SVD dimensions required to capture the network interactions, when the network is represented by a Random Dot Product Graph
(\cite{Anastasi_dimensionality_2025}). We refer to this network of connections as the "social network", meaning the network graph, rather than a "social networking site" such as Facebook. We do not suggest that there is a specific value of the embedded dimensionality at which a network becomes polarised, but rather that observation of a network over time can show polarisation occurring if the embedded dimensionality decreases over time. This can also work as a method for detecting populism if it is based on creating a deep divide between the populist group and an out-group; the political science work we draw from refers to polarisation as "populist antagonism" (\cite{laclau_hegemony_1985}). When expressed as a Random Dot Product Graph, a polarised network will have a low $\hat{d}$, and $\hat{d}$ will increase as connections are made between the groups. In the abstract, a "maximally polarised" social network would be one which is split into two groups with no connections between them at all, giving a $\hat{d}$ of 2. In real life this is not achievable due to the complexity of nation-sized social systems, but we believe that it is still possible for a nation to pursue a minimal $\hat{d}$ through systems of law and violence that are intended to create two separate poles. Examples of national systems we would consider polarised are  South Africa under apartheid and the USA during the early 20th Century era of racial segregation. 

We have turned to Schelling segregation as a straightforward way of investigating some of the dynamics underlying polarisation. If a polarised society is one where members of the two groups are prevented from making connections, then the dynamics of agents wanting certain numbers of same-group connections are likely to be meaningful for whether a network is polarised or not. The Schelling segregation model was established in 1971, with simple rules dictating agents' behaviour on a grid (\cite{schelling_segregation_1971}). Agents were "happy" when a sufficient number of their neighbours were in the same group; if not, they moved to a vacant space. Schelling found that even low levels of same-neighbour preferences resulted in patterns of segregation. Using a Schelling segregation model in a social network gives us the opportunity to further test some hypotheses about polarisation which we have already explored using Stochastic Block Models (\cite{Anastasi_dimensionality_2025}). The first hypothesis is that polarisation decreases when the groups are more connected, and the second is that polarisation decreases as one group becomes larger than the other. We found clear support for the first, and mixed support for the second; while polarisation decreased as one group became large, when it was much larger than the other (99-to-1) polarisation rapidly increased again. In a Schelling model, the critical parameter at play in for the first hypothesis is the agents' minimum number of same-group neighbours needed to be happy, and for the second hypothesis the critical parameter is the sizes of the groups.

 Many extensions to the original model have been made.\cite{clark_residential_1991} evaluated the Schelling model using real same-group preference data from surveys in the USA, and found that while the real preferences did not match Schelling's assumptions they still generated segregation. \cite{gilbert_varieties_2002} uses variations on the Schelling model to show that higher-order effects which match residential segregation can be found from many types of models, so models of residential segregation must be validated at both the individual and macro level. \cite{zhang_dynamic_2004} found that when the initial model is segregated rather than random, white-ethnicity agents will leave neighbourhoods as black-ethnicity agents enter them, preserving the segregation even when it is not optimal for their overall preferences regarding amenities etc. \cite{fossett_overlooked_2005} found that while a 50-50 ethnic mix is often used in simulation, other ethic mixes have notable effects on the results and a 50-50 mix may not apply to many real-world scenarios. They also find that a small added preference for neighbours of the other group reduces the segregation effect compared to neutrality towards the other group. \cite{fossett_preferences_2006} raises the "paradox of weak minority preferences", where the minority group can be the drivers of segregation if they want to be "more integrated" than their presence in the city allows (e.g. wanting a 50-50 neighbourhood despite an overall 80-20 ethnic mix). This minority preference pulls other members of the minority into one specific neighbourhood to satisfy their happiness requirements, which reduces the ethnic mix everywhere else on the map; as such, existing patterns of segregation can be maintained even by the tactics used by the minority to try and protect themselves. \cite{clark_context_2008} continue this line of inquiry and find that the unwillingness of any group to be a minority in "their own neighbourhood" can maintain segregation even in the absence of animosity towards other groups, even when additional dynamics such as housing cost are considered. \cite{fagolio_segregation_2007} show that segregation occurs in several types of spaces, such as non-directed graphs, indicating that Schelling segregation is not due to the topological constraints of the grid space used in the original model. Similarly, \cite{domic_dynamics_2011} used lattices in two and three dimensions to explore further spatial elements of the model and whether thermodynamic principles of energy can be used to model the emergent segregation dynamics. \cite{radi_limitations_2015} found that limiting how many of the minority group can move into a given neighbourhood at once helped promote integration, by preventing the influx from causing major destabilisation. \cite{gandica_topology_2016} found that the topology of the space affects the early stages of the model at high tolerance values, but at lower tolerance values the segregation process is topology-independent. \cite{collard_landscape_2016} added a parameter to measure how segregated each individual was from the other group, so that more segregated agents were higher vertically in the resulting plots; examining their plots as if they were topographical maps, they found that models with low other-group tolerance were segregated into "hills and valleys" rather than the more even "ruggedness" of higher-tolerance models. \cite{jani_scarcity_2020} tested the effects of housing scarcity on Schelling segregation, finding that segregation initially increased as resources became more scarce, before finally collapsing when resources were too scarce to sustain it. \cite{ishida_fuzzy_2024} explored the role of fuzzy group membership (i.e. membership as a continuous variable rather than a discrete one) on segregation patterns; they found that less-fuzzy agents tended to be more unhappy and "wander" for longer, and when they finally settled it was often near agents of the other group (since spaces near more similar agents had already been occupied), which they suggest offers another potential mechanism for explaining ethnic conflict in a neighbourhood. \cite{blair_outside_2023} found that whether white agents departed from a neighbourhood was related to the cost of outside options, including that higher desire to be near other white agents could prevent an agent from leaving the neighbourhood as it integrates, rather than departing as is commonly expected.

We are interested in expanding Schelling segregation to a social network modelled with an RDPG. Social networks are less dependent on geographic proximity than they have been in the past, since people are able to form connections online rather than only in person, so a person’s social network may not match the level of integration in e.g. their residential neighbourhood. As such, we give particular attention to the following two papers which use a network graph in this way. \cite{Henry_network_2011} use a Markov chain to generate a stochastic sequence of graphs comprised of actors (nodes) connected by links (edges). The model evolves based on Schelling-based assessments of the attribute distance between two actors connected by a given edge, with links being terminated and rewired to prioritise connections between actors with smaller attribute distances. They use networks of 100 actors connected by 300 edges, and find that segregation emerges in the graphs in similar ways to spatial Schelling models. Similarly, \cite{Gretha_network_2018} create graphs of 1000 agents with 4 initial links, and base network evolution behaviours on a tolerance of 50\%. They find that segregation emerges in the network that is greater than the amount desired by the agents, similar to in Schelling's original results (though this may not be surprising given that this effect has been observed with tolerances as low as 37.5\%). Curiously, random behaviour in the selection of which agent to make a new link with caused cascades of dissatisfaction and more movement within the network, whereas "discriminatory" selection of an agent in the same group caused fewer cascades -- an important result for understanding how the dynamics of networks are different to those of spatial models. Based on these results we believe that there is scope for new research that uses much larger and more densely connected graphs, in order to model an agent's social life more extensively. One of our aims with this paper is to address this gap.

Our first aim with this paper is to extend network-based Schelling segregation models to much denser networks than have previously been used. This allows for more granularity in the tolerance settings, because a density of 80 edges means that each change is 1.25\% of an agent's network rather than e.g. 25\% in the case where they have 4 edges. Testing a more densely connected network is also beneficial for research into online polarisation, since internet users often interact with hundreds of people (including those who they do not directly follow, but may talk to in the replies or comments of a post from someone they do follow).

Secondly, we systematically vary both the tolerance of the two groups and their proportions. The combination of these variables has been found to be very important to the Schelling model, since it is the driver of \cite{fossett_preferences_2006}'s "paradox of weak minority preferences". While a social network is not subject to the spatial considerations that have previously applied to these results, it is still possible that this effect will occur a social network model, as minority group members may become inundated with links to majority group members, making it difficult to achieve the desired ratio of same- and different-group connections. We will test this by setting the group sizes to 15\% and 85\% and making small changes to $t$ for both groups. 

Finally, we measure $\hat{d}$ alongside the ratios of same- and different-group connections for each agent, as a further investigation of whether it is a good measurement of polarisation in social networks. Our preliminary results using Stochastic Block Models indicate that $\hat{d}$ increases when there are more connections between two groups -- i.e. the groups are less polarised -- and that it decreases sharply when one group is much larger than the other. Since existing research on the Schelling model has found that segregation and agent unhappiness are both high when one group is much larger than the other, we are interested to see what results we get by measuring $\hat{d}$ for the networks generated in our experiment. 

\section{Methods}
\subsection{Agent-based model}
We designed an agent-based model in Julia \cite{Agents.jl} \cite{julia_article} that uses the preference rules of Schelling segregation to change the networking behaviour of the agents. It takes parameters $t \in [0,1]$ for tolerance ($t_1 \in [0,1]$ and $t_2 \in [0,1]$ when the tolerance is different between the groups), i.e. the proportion of same-group neighbours needed for the agent to be happy, and $s \in [0,1]$ for the size of the smaller group. The model is initialised with 1000 agents split between 2 groups, who each make 80 random connections on a simple weighted network graph. At each step, each agent evaluates how many of its network neighbours are in the same group as it. If the proportion of same-group neighbours is equal to or above its tolerance $t$, it is happy and does nothing. If the proportion of same-group neighbours is below $t$, it breaks a connection with a neighbour of the opposite group. If at any point the number of total neighbours is lower than 40, it adds a neighbour at random. The model runs for 1000 steps or until all agents are happy. All models were repeated 100 times for each combination of parameters.

\subsection{Network modelling}
In order to measure the embedded dimensionality $\hat{d}$ we model the simple weighted network graphs as Random Dot-Product Graphs (RDPGs) \cite{CareyPriebeSurvey}. This graph embedding is chosen because its optimal embedded dimensionality is established before the analysis and is independent of the network size. For a full discussion of the use of RDPG, see \cite{Anastasi_dimensionality_2025}.

\subsection{Embedding dimension}
We define the \textit{dimension} of a social network as the \textit{optimal} choice of $d$ for the RDPG embedding of the network and denote it $\hat{d}$.

The optimal choice of $\hat{d}$ can be obtained from $\Sigma$, the sorted sequence of singular values of the network's adjacency matrix $A$. There are several methods for finding $\hat{d}$; we use the elbow method proposed in \cite{zhu2006automatic}, where one Gaussian distribution is fitted to the largest $d$ singular values and another to the smallest $K-d$. The optimal value of $\hat{d}$ is the one which maximises the sum of the log-likelihoods of these two distributions.

An RDPG where every pair of nodes have an identical and independent probability of establishing a connection will have a $\hat{d}$ of 1, since 1 dimension is required for the nodes to make connections at all. It will not have a higher dimensionality because there is nothing else affecting how the edges are distributed. As such, instead of observing a high $\hat{d}$ which decreases as the network becomes segregated, as we have found in our previous work, the network will have a $\hat{d}$ of 1 until it is segregated enough that this noticeably affects whether edges are able to form between the two groups, at which point $\hat{d}$ will increase to 2. We note this because in our previous work we have used Stochastic Block Models, which have enough structure to have a higher $\hat{d}$, whereas the random simple weighted graphs used in this paper are much more random and therefore start with a $\hat{d}$ of 1.

\subsection{Code and data}
The code used for generating, analysing, and plotting the data, as well as the data itself and the plots used in this paper, are available at https://doi.org/10.5281/zenodo.15265494 \cite{Anastasi_repository}

\section{Results}
\subsection{Symmetrical tolerances}
We set $s=0.5$, giving 500 agents of each group in the network for a total of 1000 agents. We varied $t$ from 0.05 to 1 in increments of 0.05. We measured the number of steps it took for all agents to become happy (fig1), the proportion of similar neighbours to total neighbours (fig2), and $\hat{d}$, with full results reported in Table 1.

\begin{table}[htb!]
\centering
\begin{tabular}{rrrr}
  \hline
Tolerance & Similarity & Dimension & Stabilisation \\ 
  \hline
0.05 & 0.50 & 1.00 & 1.00 \\ 
0.10 & 0.50 & 1.00 & 1.00 \\ 
0.15 & 0.50 & 1.00 & 1.00 \\ 
0.20 & 0.50 & 1.00 & 1.00 \\ 
0.25 & 0.50 & 1.00 & 1.00 \\ 
0.30 & 0.50 & 1.00 & 1.00 \\ 
0.35 & 0.50 & 1.00 & 1.07 \\ 
0.40 & 0.50 & 1.00 & 10.30 \\ 
0.45 & 0.50 & 1.00 & 24.34 \\ 
0.50 & 0.52 & 1.00 & 33.45 \\ 
0.55 & 0.57 & 1.00 & 38.53 \\ 
0.60 & 0.62 & 1.00 & 41.45 \\ 
0.65 & 0.67 & 1.00 & 43.40 \\ 
0.70 & 0.72 & 1.00 & 45.01 \\ 
0.75 & 0.77 & 2.00 & 45.97 \\ 
0.80 & 0.82 & 2.00 & 47.03 \\ 
0.85 & 0.87 & 2.00 & 47.57 \\ 
0.90 & 0.92 & 2.00 & 47.24 \\ 
0.95 & 0.97 & 2.00 & 46.35 \\ 
1.00 & 1.00 & 2.00 & 40.78 \\ 
   \hline

\end{tabular}
\caption{Table of results for same-size groups with symmetrical tolerances.} 
\label{Table 1}
\end{table}

\begin{figure} [hbt!]
\includegraphics[width=\textwidth]{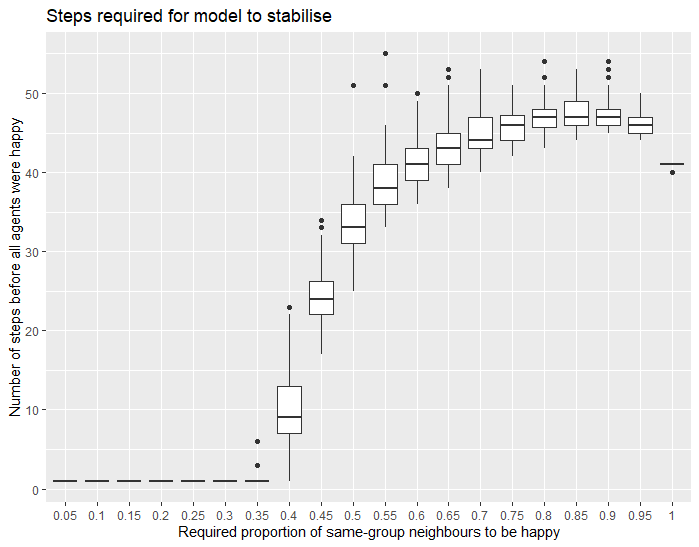}
\caption{Boxplot of how many steps each model took for all agents to become happy, at each tolerance level.} \label{fig1}
\end{figure} 

For tolerances $t \le 0.3$ the network was immediately stable. This is because in the case of $s=0.5$ group size and random neighbour assignment, it is simply very unlikely that any agent would have fewer than 30\% of its neighbours in the same group. We tested $t=1$ (no same-group neighbours) to find the baseline time for the network to stabilise and found that this was 40.8 steps. For tolerances where stabilisation took longer than this, $[0.6 \le t \le 0.95]$, it is likely that the random behaviour settings for adding new friends meant that some agents were severing connections with the other group and then immediately making a new connection with it. However, we do not seem to see the "cascades of dissatisfaction" found by \cite{Gretha_network_2018} in their model with random behaviour, since the model was still able to stabilise in under 50 steps. Finally, for $0.35 \le t \le 0.55$, the network did not stabilise immediately, but did stabilise faster than for $t = 1$. 

\begin{figure} [hbt!]
\includegraphics[width=\textwidth]{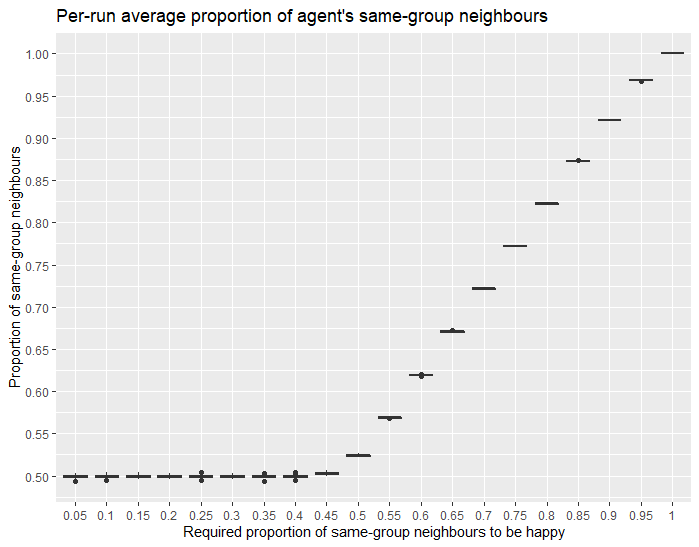}
\caption{Boxplot of the average proportion of each agent's neighbours who are in the same group as it, at each tolerance level.} \label{fig2}
\end{figure} 

The similarity ratio for the network when $t \le 0.45$ was 0.5. This is due to the group size of $s=0.5$, since this is the similarity ratio when the model is initialised and the agents are immediately happy, so they do not take action to reduce the similarity of their neighbours. For tolerances $[0.5 \le t \le 0.95]$ the similarity ratio was 0.02 greater than the tolerance, i.e an agent with a 50\% tolerance had 52\% similar neighbours. This indicates that the network was able to reach an arrangement where all the agents were happy without exceeding the tolerance ratio, which is different to many Schelling models where the similarity ratio becomes much greater than the agents' tolerance. The high density of the network means that as other-group connections are removed, the similarity ratio increases slowly; this means that the agent does not "overshoot" their tolerance by a large amount.

The embedded dimensionality of the network was 1 for $t \le 0.7$ and 2 for $t \ge 0.75$. Under our definition this indicates that the network was segregated enough to be "polarised" when the agents' happiness required at least 75\% same-group neighbours. Since there was no point at which the similarity ratio began to greatly exceed the tolerance, we take $\hat{d}$ as the "tipping point" for our models. It is notable that the tipping point for a densely connected network graph seems to be much higher than for sparsely connected graphs or space-based Schelling models.

\subsection{Asymmetrical tolerances}
An important update to the original Schelling model was the incorporation of asymmetrical tolerances, based on the observed preferences of ethnic groups in cities in the USA. This dynamic is important because one group's intolerance may be enough to create segregation by itself, meaning that no amount of intolerance from the other group can overcome the divide. To test how this applies to a densely connected social network, we retained the $s=0.5$ group sizes from the previous model, fixed group $t_1$ to either 0.5 (which does not produce segregation) or 0.9 (which does), and varied $t_2$ from 0.05 to 1 in steps of 0.05. We measured the number of steps it took for all agents to become happy (fig3), the proportion of similar neighbours to total neighbours (fig4), and $\hat{d}$ (fig5), with full results reported in Table 2.

\begin{table}[hbt!]
\centering
\begin{tabular}{rrrrrrr}
  \hline
$t_1$ & $t_2$ & Similarity & Stabilisation & G1 Similarity & G2 Similarity & Dimension \\ 
  \hline
0.50 & 0.05 & 0.52 & 35.90 & 0.51 & 0.52 & 1.00 \\ 
0.50 & 0.10 & 0.52 & 35.28 & 0.51 & 0.52 & 1.00 \\ 
0.50 & 0.15 & 0.52 & 35.35 & 0.51 & 0.52 & 1.00 \\ 
0.50 & 0.20 & 0.52 & 35.81 & 0.51 & 0.52 & 1.00 \\ 
0.50 & 0.25 & 0.52 & 35.23 & 0.51 & 0.52 & 1.00 \\ 
0.50 & 0.30 & 0.52 & 35.84 & 0.51 & 0.52 & 1.00 \\ 
0.50 & 0.35 & 0.52 & 35.60 & 0.51 & 0.52 & 1.00 \\ 
0.50 & 0.40 & 0.52 & 35.42 & 0.51 & 0.52 & 1.00 \\ 
0.50 & 0.45 & 0.52 & 35.49 & 0.51 & 0.52 & 1.00 \\ 
0.50 & 0.50 & 0.52 & 34.29 & 0.52 & 0.53 & 1.00 \\ 
0.50 & 0.55 & 0.55 & 43.42 & 0.55 & 0.56 & 1.00 \\ 
0.50 & 0.60 & 0.60 & 52.22 & 0.60 & 0.60 & 1.00 \\ 
0.50 & 0.65 & 0.65 & 60.36 & 0.65 & 0.65 & 1.00 \\ 
0.50 & 0.70 & 0.70 & 66.50 & 0.70 & 0.70 & 1.00 \\ 
0.50 & 0.75 & 0.75 & 73.53 & 0.75 & 0.75 & 1.76 \\ 
0.50 & 0.80 & 0.80 & 78.07 & 0.80 & 0.80 & 2.00 \\ 
0.50 & 0.85 & 0.85 & 82.54 & 0.85 & 0.86 & 2.00 \\ 
0.50 & 0.90 & 0.90 & 87.15 & 0.90 & 0.91 & 2.00 \\ 
0.50 & 0.95 & 0.96 & 91.14 & 0.95 & 0.96 & 2.00 \\ 
0.50 & 1.00 & 1.00 & 93.96 & 1.00 & 1.00 & 2.00 \\ 
0.90 & 0.05 & 0.91 & 90.34 & 0.90 & 0.91 & 2.00 \\ 
0.90 & 0.10 & 0.91 & 90.50 & 0.90 & 0.91 & 2.00 \\ 
0.90 & 0.15 & 0.91 & 90.21 & 0.90 & 0.91 & 2.00 \\ 
0.90 & 0.20 & 0.91 & 90.05 & 0.90 & 0.91 & 2.00 \\ 
0.90 & 0.25 & 0.91 & 90.41 & 0.90 & 0.91 & 2.00 \\ 
0.90 & 0.30 & 0.91 & 90.08 & 0.90 & 0.91 & 2.00 \\ 
0.90 & 0.35 & 0.91 & 89.69 & 0.90 & 0.91 & 2.00 \\ 
0.90 & 0.40 & 0.91 & 89.59 & 0.90 & 0.91 & 2.00 \\ 
0.90 & 0.45 & 0.91 & 89.40 & 0.90 & 0.91 & 2.00 \\ 
0.90 & 0.50 & 0.91 & 87.86 & 0.90 & 0.91 & 2.00 \\ 
0.90 & 0.55 & 0.91 & 82.14 & 0.90 & 0.91 & 2.00 \\ 
0.90 & 0.60 & 0.91 & 76.56 & 0.90 & 0.91 & 2.00 \\ 
0.90 & 0.65 & 0.91 & 70.68 & 0.90 & 0.91 & 2.00 \\ 
0.90 & 0.70 & 0.91 & 65.81 & 0.90 & 0.91 & 2.00 \\ 
0.90 & 0.75 & 0.91 & 60.52 & 0.90 & 0.91 & 2.00 \\ 
0.90 & 0.80 & 0.91 & 56.77 & 0.90 & 0.91 & 2.00 \\ 
0.90 & 0.85 & 0.91 & 51.82 & 0.91 & 0.91 & 2.00 \\ 
0.90 & 0.90 & 0.92 & 47.16 & 0.92 & 0.92 & 2.00 \\ 
0.90 & 0.95 & 0.96 & 51.37 & 0.96 & 0.96 & 2.00 \\ 
0.90 & 1.00 & 1.00 & 54.62 & 1.00 & 1.00 & 2.00 \\ 
   \hline
\end{tabular}
\caption{Table of results for same-size groups with asymmetrical tolerances.} \ \label{Table 2}
\end{table}

\begin{figure} [hbt!]
\includegraphics[width=\textwidth]{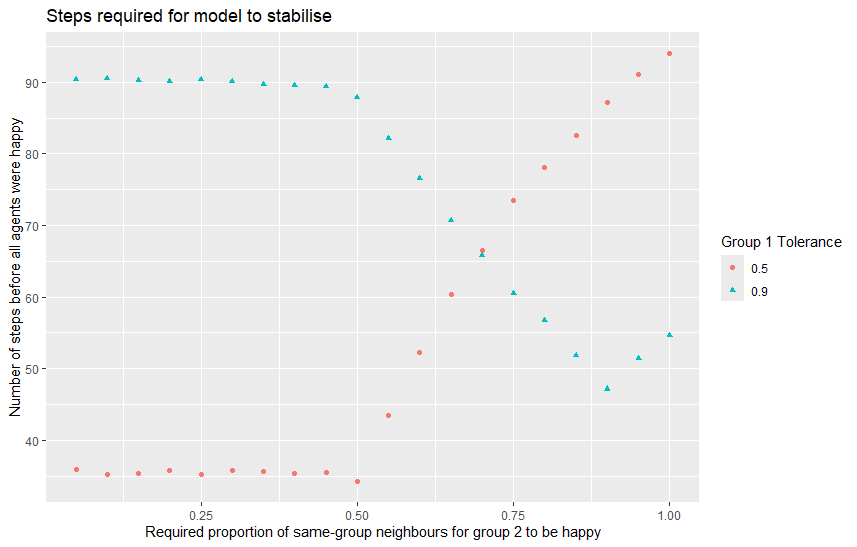}
\caption{Plot of the average number of steps required for every agent to become happy.} \label{fig3}
\end{figure} 

When $t_1 = 0.5$, the model stabilised in under 40 steps if $t_2 \le 0.5$. As $t_2$ increased beyond 0.5 took longer to stabilise, peaking at 94 steps when $t_2 = 1$. When $t_1 = 0.9$, the model took 89 or 90 steps to stabilise if the second group's tolerance was below 0.5, and then began to stabilise more rapidly as the second group required more same-group neighbours to be happy. Notably, in this case the model stabilised the fastest when both groups were set to 0.9 and then began to take longer again when set to 0.95 or 1.

\begin{figure} [hbt!]
\includegraphics[width=\textwidth]{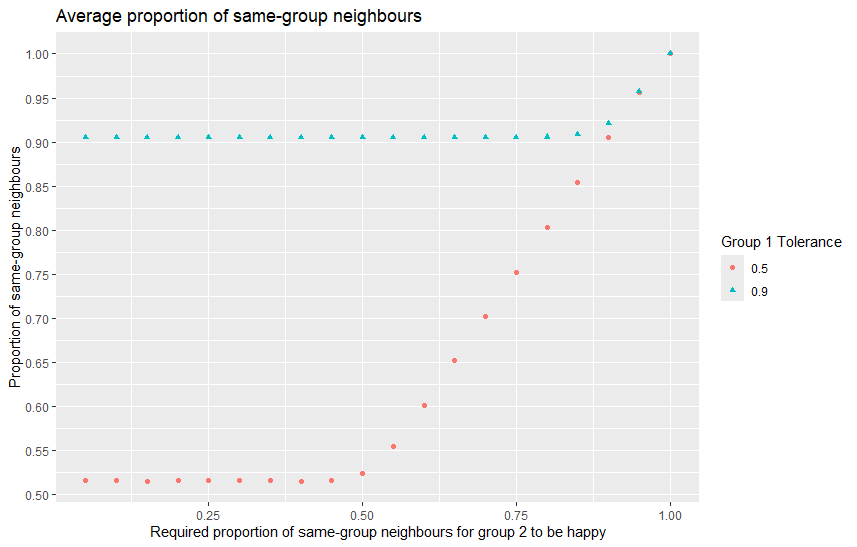}
\caption{Plot of the average proportion of each agent's neighbours who are in the same group as it, at each tolerance level.} \label{fig4}
\end{figure} 

The behaviour of the similarity ratio was strongly affected by the fixed tolerance $t_1$e. When this was set to 0.9, both groups had a similarity ratio of 0.9, no matter how low $t_2$ was. When considering the behaviour of the network, this makes intuitive sense; group 1 is limiting the number of connections it will make to group 2, so group 2 cannot exceed that number of connections even if it would be happy to do so. When $t_1$ was fixed at 0.5, the similarity ratio was 0.52 for $t \le 0.5$, and it tracked the change in $t_2$ for $t_2 \ge 0.55$, . This reflects the fact that $t_1$ set a floor for the similarity ratio, as we saw when testing symmetrical tolerances. For this case the differences in similarity ratio between group 1 and 2 were negligible, so they are recorded but not plotted.

\begin{figure} [hbt!]
\includegraphics[width=\textwidth]{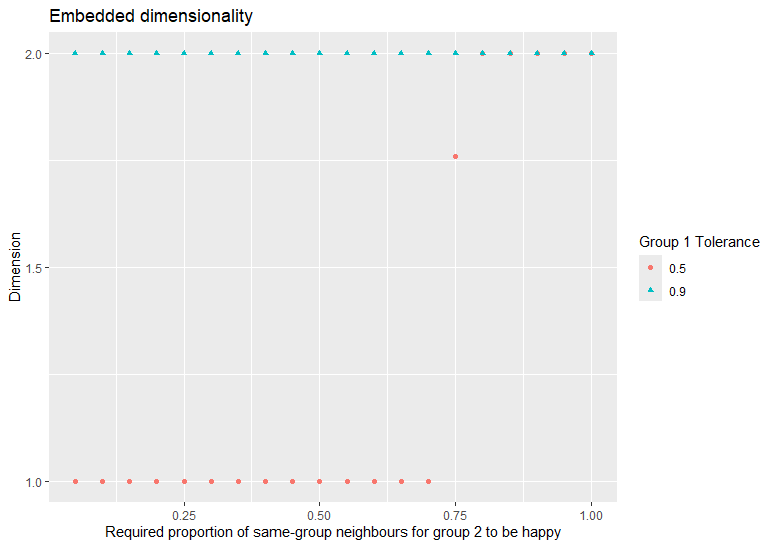}
\caption{Plot of the average embedded dimension of the networks.} \label{fig5}
\end{figure} 

The embedded dimensionality $\hat{d}$ of the model was 1 (non-polarised) when both groups had $t \le 0.7$ and 2 (polarised) when either group had $t \ge 0.8$. Interestingly, when one group had $t = 0.5$ and the other had $t = 0.75$, $\hat{d}$ was 1 in $24 / 100$ runs and 2 in all others, resulting in an average embedded dimensionality of 1.76. This contrasts with our earlier result where all runs were polarised when the groups had symmertical $t= 0.75$, suggesting that greater tolerance from one of the groups may have a small effect on whether the network becomes polarised.

\subsection{Changing group size}
To investigate the effect of changing group size, we ran the same model with symmetrical tolerances of $t=0.25$, $t=0.5$, $t=0.85$, and $t=0.95$ so that group size effects were captured when the model was polarised and when it was not. For each tolerance, a minority group size $0.05 \le s \le 0.5$ was tested in increments of 0.05. We could not test $s\le0.05$ due to the minimum of 40 neighbours set for the agent; since $s=0.01$ would give only 10 agents in the minority, they would be structurally incapable of being happy if they have $t_1 \ge 0.25$ and the model would be unable to stabilise at all. We measured the number of steps it took for all agents to become happy (fig6), the proportion of similar neighbours to total neighbours (fig7), and $\hat{d}$, with full results reported in Table 3.

\begin{table}[hbt!]
\centering
\begin{tabular}{rrrrrrr}
  \hline
Size & Tolerance & Similarity & Stabilisation & Group 1 Similarity & Group 2 Similarity & Dimension \\ 
  \hline
0.05 & 0.25 & 0.95 & 319.61 & 0.28 & 0.99 & 1.00 \\ 
0.05 & 0.50 & 0.97 & 562.86 & 0.52 & 0.99 & 1.00 \\ 
0.05 & 0.85 & 0.99 & 953.46 & 0.86 & 1.00 & 1.00 \\ 
0.10 & 0.25 & 0.89 & 161.96 & 0.25 & 0.96 & 1.00 \\ 
0.10 & 0.50 & 0.94 & 278.74 & 0.52 & 0.98 & 1.00 \\ 
0.10 & 0.85 & 0.98 & 413.32 & 0.86 & 1.00 & 1.00 \\ 
0.10 & 0.95 & 0.99 & 402.02 & 0.95 & 1.00 & 1.00 \\ 
0.15 & 0.25 & 0.82 & 102.91 & 0.25 & 0.92 & 1.00 \\ 
0.15 & 0.50 & 0.90 & 177.83 & 0.51 & 0.97 & 1.00 \\ 
0.15 & 0.85 & 0.97 & 264.46 & 0.86 & 0.99 & 1.00 \\ 
0.15 & 0.95 & 0.99 & 221.91 & 0.95 & 1.00 & 1.00 \\ 
0.20 & 0.25 & 0.72 & 79.59 & 0.25 & 0.84 & 1.00 \\ 
0.20 & 0.50 & 0.85 & 128.24 & 0.50 & 0.94 & 1.00 \\ 
0.20 & 0.85 & 0.96 & 179.05 & 0.86 & 0.99 & 1.00 \\ 
0.20 & 0.95 & 0.99 & 147.73 & 0.95 & 1.00 & 1.00 \\ 
0.25 & 0.25 & 0.64 & 54.17 & 0.26 & 0.76 & 1.00 \\ 
0.25 & 0.50 & 0.80 & 105.39 & 0.50 & 0.90 & 1.00 \\ 
0.25 & 0.85 & 0.95 & 119.27 & 0.86 & 0.98 & 1.00 \\ 
0.25 & 0.95 & 0.98 & 104.50 & 0.95 & 0.99 & 1.00 \\ 
0.30 & 0.25 & 0.58 & 29.83 & 0.30 & 0.70 & 1.00 \\ 
0.30 & 0.50 & 0.74 & 90.87 & 0.50 & 0.84 & 1.00 \\ 
0.30 & 0.85 & 0.94 & 81.41 & 0.86 & 0.97 & 1.00 \\ 
0.30 & 0.95 & 0.98 & 73.28 & 0.96 & 0.99 & 1.00 \\ 
0.35 & 0.25 & 0.54 & 5.62 & 0.35 & 0.65 & 1.00 \\ 
0.35 & 0.50 & 0.68 & 76.86 & 0.50 & 0.78 & 1.00 \\ 
0.35 & 0.85 & 0.92 & 64.35 & 0.86 & 0.95 & 2.00 \\ 
0.35 & 0.95 & 0.98 & 54.81 & 0.96 & 0.99 & 2.00 \\ 
0.40 & 0.25 & 0.52 & 1.00 & 0.40 & 0.60 & 1.00 \\ 
0.40 & 0.50 & 0.62 & 63.79 & 0.50 & 0.69 & 1.00 \\ 
0.40 & 0.85 & 0.90 & 57.94 & 0.86 & 0.93 & 2.00 \\ 
0.40 & 0.95 & 0.97 & 48.38 & 0.96 & 0.98 & 2.00 \\ 
0.45 & 0.25 & 0.50 & 1.00 & 0.45 & 0.55 & 1.00 \\ 
0.45 & 0.50 & 0.56 & 49.58 & 0.50 & 0.60 & 1.00 \\ 
0.45 & 0.85 & 0.88 & 52.34 & 0.86 & 0.90 & 2.00 \\ 
0.45 & 0.95 & 0.97 & 47.36 & 0.96 & 0.97 & 2.00 \\ 
0.50 & 0.25 & 0.50 & 1.00 & 0.50 & 0.50 & 1.00 \\ 
0.50 & 0.50 & 0.52 & 34.29 & 0.52 & 0.52 & 1.00 \\ 
0.50 & 0.85 & 0.87 & 47.67 & 0.87 & 0.87 & 2.00 \\ 
0.50 & 0.95 & 0.97 & 46.19 & 0.97 & 0.97 & 2.00 \\ 
   \hline
\end{tabular}
\caption{Table of results when one of the groups was smaller than the others.} \label{Table 3}
\end{table}

\begin{figure} [hbt!]
\includegraphics[width=\textwidth]{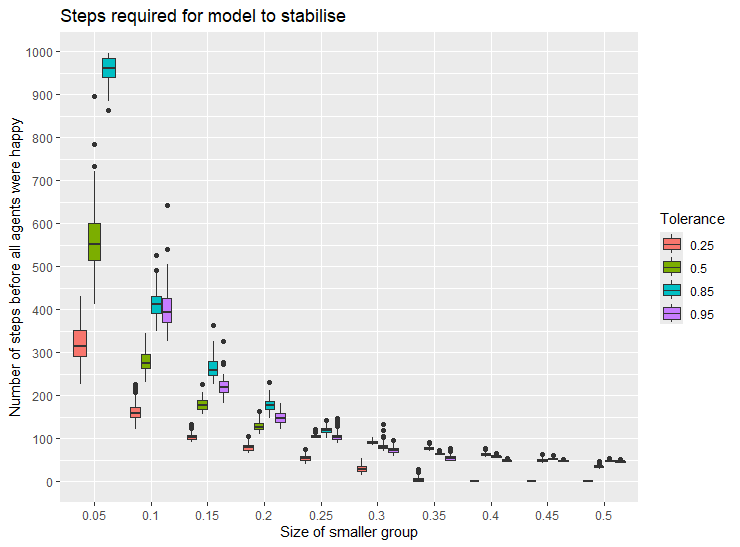}
\caption{Boxplot of the number of steps required for each agent to be happy. Note that the data for group size 0.05 at tolerance 0.95 are omitted, since these models did not stabilise, and group size 0.05 at tolerance 0.85 is based on a smaller sample size than other boxes.} \label{fig6}
\end{figure} 

The model was unable to stabilise when $s = 0.05$ and $t = 0.95$. It also struggled to stabilise when $s = 0.05$ and $t = 0.85$, with only 24 out of 100 runs stabilising and all of them taking more than 850 steps. This is noteworthy because the small group in this model had 50 agents, so it was technically possible for these agents to be happy at $t=0.95$ if they gained 38 same-group neighbours, but this did not occur. We suspect this was due to the very low chance of these agents finding each other by making new friends at random, as well as the high chance of opposite-group agents making unwanted connections that they must spend more time removing. At this group size the network stabilised the fastest with a tolerance of 0.25, and slowest with a tolerance of 0.85. For $[0.1 \le s \le 0.25]$ the model stabilised slowest with $t=0.85$ and fastest with $t=0.25$. For $[0.3 \le s \le 0.35]$, the model stabilises slowest with $t = 0.5$ and fastest with $t = 0.25$. For $s \ge 0.4$ the network stabilised immediately for $t = 0.25$, and had negligible differences in stabilisation speed for all other $t$ (i.e. for each group size $s$ the stabilisation speed was different, but for a given $s$ the stabilisation speed was only different for $t=0.25$ and was the same for all other $t$). Overall the stabilisation speed shows a consistent pattern with very high or low tolerances, but medium tolerances behave differently at different group sizes.

\begin{figure} [hbt!]
\includegraphics[width=\textwidth]{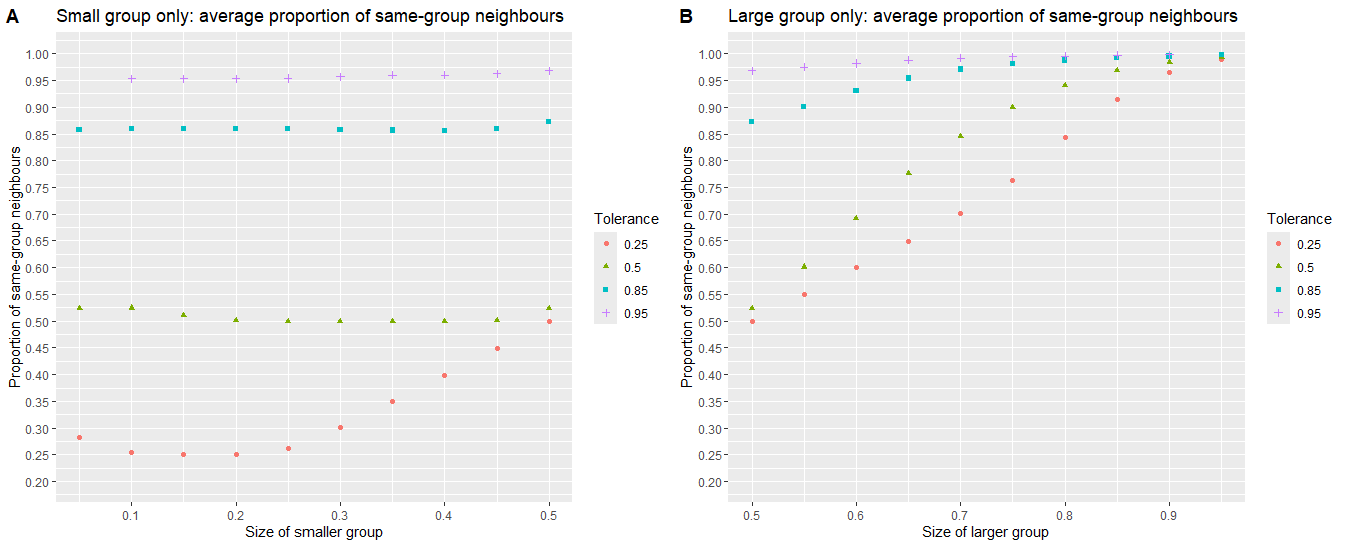}
\caption{Plot of the average proportion of each agent's neighbours who are in the same group as it, at each tolerance level.} \label{fig7}
\end{figure} 

The difference between the similarity ratio and the tolerance was greatly affected by the group size. For example, in the case where $t = 0.25$ and $s = 0.05$, the difference was 0.7, meaning that an agent had an average of 95\% same-group neighbours despite being happy with only 25\%. When splitting this up by which group the agent belonged to, we find that the large group had a similarity of 0.99, whereas the smaller group had a similarity of 0.28. For $t = 0.25$ the smaller group's similarity increased with its group size, reflecting previous results about similarity exceeding tolerance when $t < s$. 

As expected, the networks with $t = 0.25$ or $t=0.5$ did not become polarised for any value of $s$. For $t=0.85$ and $t=0.95$, the network became polarised when $s \ge 0.35$. Thus, it could be said that the polarisation "tipping point" is only present when the groups are close to being the same size, and is not an issue when the difference between the group sizes is greater.

This result is also interesting because it augments our previous work with stochastic block models, which found that polarisation decreased as the group size difference increased, except for a sharp increase in polarisation when the group sizes were 99\textendash1 (i.e $s=0.01$). We were unable to fully test the $s=0.01$ case here because the model was structurally incapable of stabilising. However, since instability has been noted as a factor in conflicts between groups (\cite{clark_residential_1991}\cite{fossett_overlooked_2005},\cite{fossett_preferences_2006},\cite{MACY01062006}), cases where the model could not stabilise (the $s=0.01$ case, and the $s=0.05$ case when $t = 0.95$ or $t =0.85$) could be worth investigating more closely in future research.

\subsection{The paradox of weak minority preferences}
An effect established in spatial versions of the Schelling model, but not yet tested in social network versions, is the "paradox of weak minority preferences". This is the case where the smaller group in an unequal divide wishes to have more neighbours than their overall group size, such as a group that is 15\% of the agents wishing to have 50\% same-group neighbours. In a spatial model, this causes minority agents to move away from neighbourhoods where they have 15\% same-group neighbours and towards neighbourhoods where they have more, leaving those original neighbourhoods with very low percentages of minority agents or none at all. However, it is not clear how this would apply to a densely connected social network, since many of the constraints of the spatial network are diminished.

Following \cite{MACY01062006} and \cite{fossett_preferences_2006}, we set $s=0.15$. We tested tolerance combinations of $t_1=0.5, t_2=0.5$; $t_1=0.5,t_2=0.9$; $t_1=0.9,t_2=0.5$; and $t_1=0.9,t_2=0.9$ to investigate the effects of changing tolerance on each group. Full results are given in Table 4. Further, following \cite{fossett_preferences_2006} we also tested a 5\% increase in desired same-group neighbours using tolerance combinations of $t_1=0.15,t_2=0.85$; $t_1=0.15,t_2=0.9$; $t_1=0.2,t_2=0.85$; and $t_1=0.2,t_2=0.9$ in order to explore whether a small change in the minority's preferences was more impactful than a small change in the majority's. Full results are given in Table 5. Our model does not include a preference to be connected to the other group. 

As was discovered in the tests of group size, $s=0.15$ results in a $\hat{d}$ of 1 regardless of the tolerance levels of the groups. This result occurred in this test as well, so it is not graphed or discussed further.

\subsubsection{Large changes in preference}
\begin{table}[hbt!]
\centering
\begin{tabular}{rrrrrr}
  \hline
$t_1$ & $t_2$ & Similarity & Stabilisation & G1 Similarity & G2 Similarity \\ 
  \hline
0.50 & 0.50 & 0.90 & 185.18 & 0.51 & 0.97 \\ 
0.50 & 0.90 & 0.90 & 136.61 & 0.51 & 0.97 \\ 
0.90 & 0.50 & 0.98 & 291.13 & 0.91 & 0.99 \\ 
0.90 & 0.90 & 0.98 & 246.18 & 0.91 & 0.99 \\ 
   \hline
\end{tabular}
\caption{Table of results when one of the groups was smaller than the others, .} \label{Table 4}
\end{table}

\begin{figure} [hbt!]
\includegraphics[width=\textwidth]{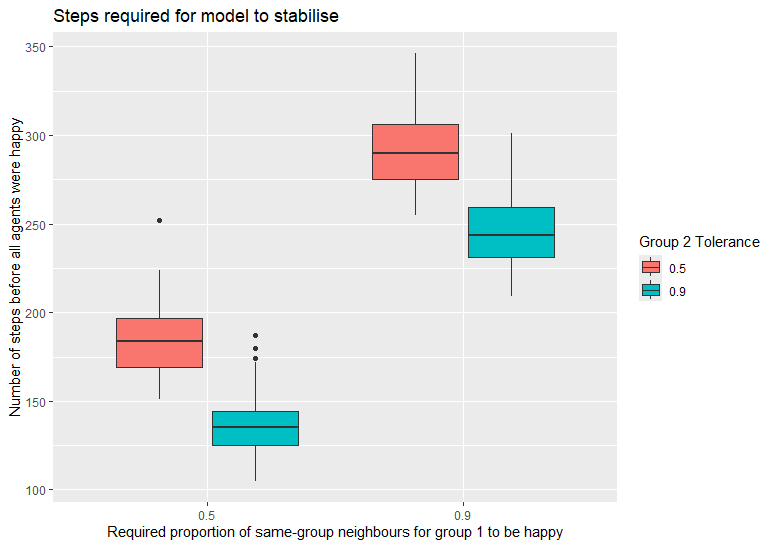}
\caption{Boxplot of the number of steps required before all agents were happy.} \label{fig8}
\end{figure} 

The fastest combination to stabilise (fig8) was $t_1=0.5,t_2=0.9$, in an average of 137 steps. The symmetrical model with $t=0.5$ stabilised in 185 steps, the symmetrical model with $t=0.9$ stabilised in 246 steps, and the $t_1=0.9,t_2=0.5$ model took the longest to stabilise at 291 steps. Both models where $t_1=0.5$ stabilised faster than the models with $t_1=0.9$ This suggests that the small group having high intolerance may make it more difficult for the model to stabilise, possibly because they only find other members of their group through random chance and it takes longer for them to find enough members of their own group.

When $t_1 = 0.5$ the increase in $t_2$ from 0.5 to 0.9 resulted in the model taking 48 fewer steps to stabilise. Similarly, when $t_1=0.9$, the increase in $t_2$ resulted in the model taking 45 fewer steps to stabilise. When $t_2=0.5$, the increase in $t_1$ from 0.5 to 0.9 resulted in the model taking 106 more steps to stabilise. Finally, when $t_2=0.9$ the increase in $t_1$ resulted in the model taking 106 steps more to stabilise. As such, changes in $t_1$ had a greater effect on how long the model took to stabilise, with changes in $t_2$ having smaller effects. However it is worth that changes in $t_1$ and $t_2$ had opposite effects, with $t_1$ increasing the steps required to stabilise and $t_2$ decreasing them. This arguably supports the claim that the preferences of the minority may increase segregation in the network.

\begin{figure} [hbt!]
\includegraphics[width=\textwidth]{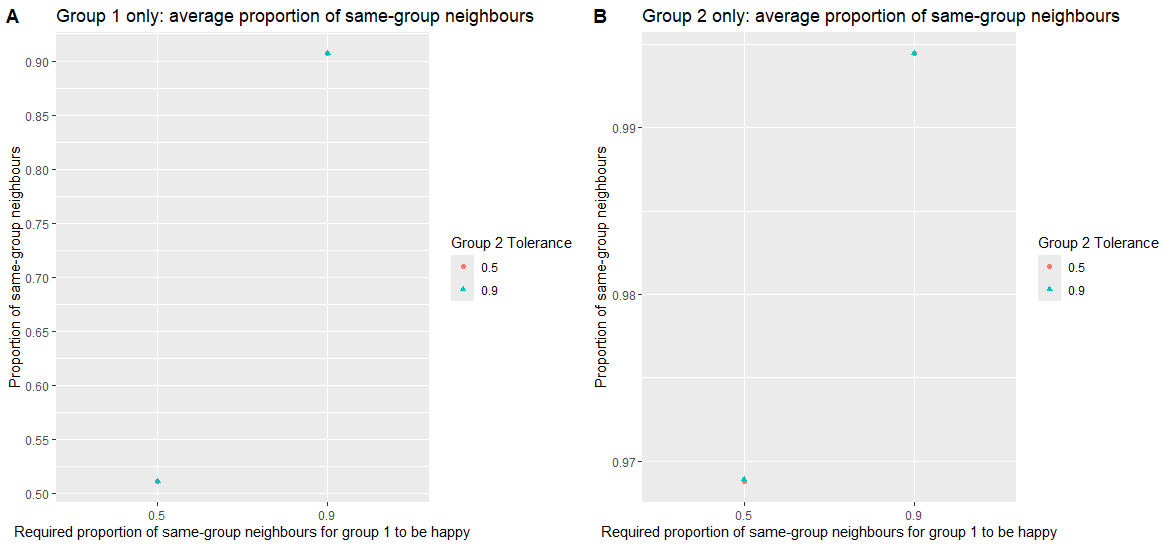}
\caption{Plot of the average number of same-group neighbours for each agent.} \label{fig9}
\end{figure} 

When the $t_1=0.5$, the similarity ratio of group 1 was 0.51 and group 2 was 0.97 (fig9). This was the case when both $t_2=0.5$ and $t_2=0.9$, which indicates that the similarity ratio is driven by $t_1$ rather than $t_2$. Notably, group 2 has a high similarity even with $t_1=0.5,t_2 =0.5$, which is likely to be a result of its larger size as previously established in the testing of group size.

In contrast, when $t_1=0.9$, the similarity ratio of group 1 was 0.90 and group 2 was 0.99. Again, this was the case when both $t_2=0.5$ and $t_2=0.9$. This indicates that the similarity of group 1 is driven by $t_1$, while $t_2$ had little influence on the similarity of group 2 and group 2's similarity is most likely to be driven by its size. 

\subsubsection{Small changes in preference}
\begin{table}[hbt!]
\centering
\begin{tabular}{rrrrrr}
  \hline
$t_1$ & $t_2$ & Similarity & Stabilisation & G1 Similarity & G2 Similarity \\ 
  \hline
0.15 & 0.85 & 0.76 & 56.90 & 0.17 & 0.87 \\ 
0.15 & 0.90 & 0.80 & 27.18 & 0.22 & 0.90 \\ 
0.20 & 0.85 & 0.79 & 77.51 & 0.20 & 0.89 \\ 
0.20 & 0.90 & 0.81 & 44.47 & 0.23 & 0.91 \\ 
   \hline
\end{tabular}
\caption{Table of results when one of the groups was smaller than the others.} \label{Table 5}
\end{table} 

\begin{figure} [hbt!]
\includegraphics[width=\textwidth]{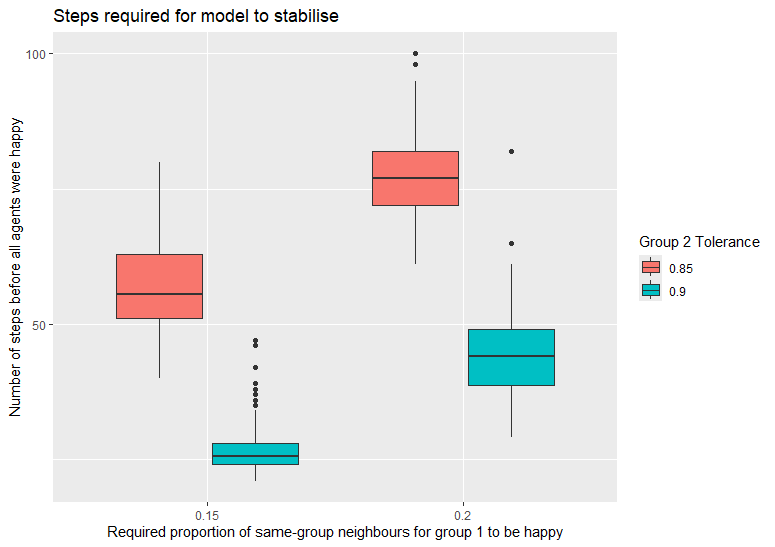}
\caption{Boxplot of the number of steps required before all agents were happy.} \label{fig10}
\end{figure} 

The fastest combination to stabilise was $t_1=0.15,t_2=0.9$, in an average of 27 steps (fig10). The second fastest was $t_1=0.2,t_2=0.9$ in 45 steps, the third fastest was $t_1=0.15,t_2=0.85$ in 57 steps, and the slowest was $t_1=0.2,t_2=0.85$ in 76 steps. Both models where $t_2 = 0.9$ stabilised faster than either of the ones where $t_2=0.85$. When $t_1=0.2$ the increase in $t_2$ from 0.85 to 0.9 resulted in the model taking 31 steps fewer to stabilise, and when $t_1=0.15$ the increase in $t_2$ resulted in it taking 30 fewer steps to stabilise. On the other hand, when $t_2=0.85$ the increase from $t_1=0.15$ to $t_1=0.2$ resulted in it requiring 19 more steps to stabilise, and when $t_2=0.9$ the increase in $t_1$ gave an increase of 18 steps. This suggests that in this case, the stabilisation speed was more driven by $t_2$ than by $t_1$, and we again observed that changing $t_1$ made the network stabilise slower and changing $t_2$ made it stabilise faster.

\begin{figure} [hbt!]
\includegraphics[width=\textwidth]{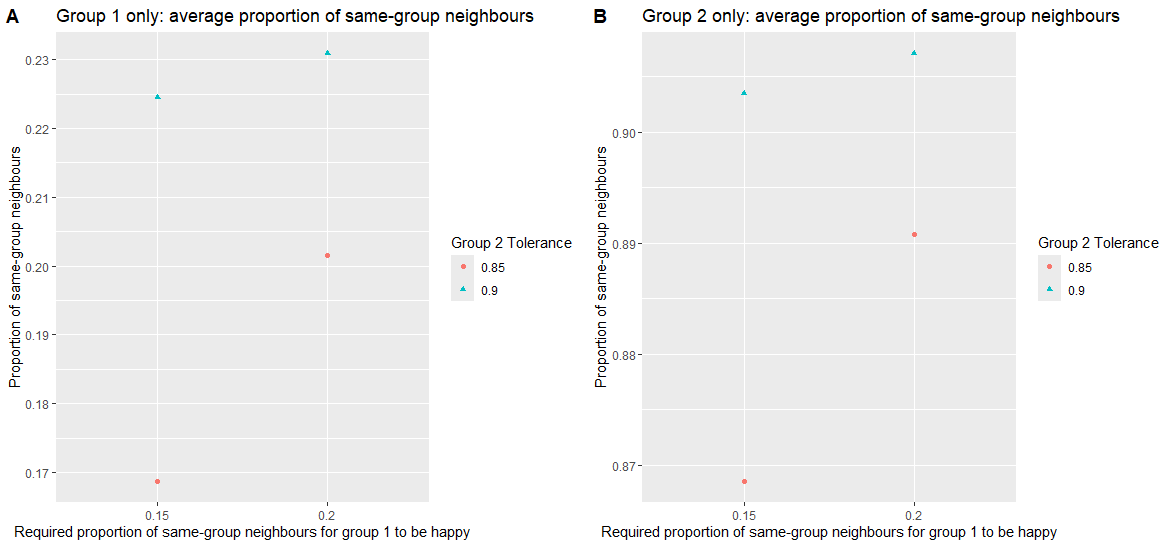}
\caption{Plot of the average number of same-group neighbours for each agent.} \label{fig11}
\end{figure} 

When the $t_2= 0.85$, the similarity ratios for both groups closely matched the tolerances of the groups (fig11). However, when the $t_2= 0.9$, the similarity ratio of group 1 (the small group) was higher than would be expected by this close matching; the small group had a similarity ratio of 0.23 even though $t_1=0.15$. The result for the $t_1=0.15$ models are the most remarkable, since an increase of $t_2$ from 0.85 to 0.9 gives a similarity increase in group 1 of 0.05, i.e. an increase in $t_2$ increases the similarity of the neighbours of group 1 (the small group), even if the minority group's tolerance $t_1$ does not change. We also note that this increase is higher than the amount we have observed tolerances to "overshoot" by, which is 0.02 or 0.03. This runs counter to the observation by \cite{fossett_preferences_2006} that the behaviour of the minority group is more separation-producing that the majority group.

\section{Conclusions}
We have found that Schelling segregation dynamics are quite different in a densely connected network than in sparsely connected networks. In a dense network, the number of similar neighbours closely matches the agents' desired number, rather than becoming much larger as is often observed in Schelling segregation. The number of same-group neighbours required to create polarisation (as measured using our embedded dimensionality method) was also much higher than previously observed tipping points in Schelling models. This may have implications for research on polarisation in social networks, such as online social media, as these networks may be structurally harder to polarise than sparse networks or residential neighbourhoods.

Investigation of asymmetric group tolerances indicated that the tipping point may be slightly higher if one group is more tolerant than the other, though polarisation will still occur if one group is very intolerant. This is relevant to other research because it is widely agreed that different groups have different tolerance levels, and so modelling the effects of this difference is important to capture the dynamics of polarisation.

Group size seemed to have some interactions with tolerance that could be worth further investigation. We found that when one group was very small, it was difficult or impossible for the model to stabilise. Even when it was mathematically possible for the model to stabilise, such as when the s$s=0.05$, high required numbers of same-group neighbours made the model struggle to stabilise or be incapable of doing so. However, when the group was small but not very small, we found that the network could not be polarised even at tolerance levels which polarised the $s=0.5$ network. This matched our previous results with stochastic block models, and suggests that for these networks, individual agents' desire for high numbers of same-group neighbours may not be able to polarise these networks by itself. Note that this result should not be overstated and that these networks may still become polarised if, for example, laws are passed that force the groups to become completely separated. However the result is still interesting given questions in the field about whether desire for same-group neighbours is capable of producing polarisation by itself.

Finally, we find that in a dense social network the evidence for the "paradox of weak minority preferences" is mixed. We found that an increase in the minority preference $t_1$ was associated with the model taking longer to stabilise, while an increase in majority preference $t_2$ was associated with faster stabilisation. However, the similarity of the agents' neighbours gave different results, finding that an increase in majority preference $t_2$ was associated with the minority group's similarity increasing, i.e. the minority had more same-group neighbours even though their tolerance $t_1$ had not changed. While the time required for the model to stabilise is important, we believe that the results about the similarity are more salient, and they do not indicate that minority preference is a larger driver of similarity that majority preference. More research is needed on this particular effect.

There are many extensions that could be made to our model. It would be straightforward to add a desired number of opposite group neighbours, fuzzy set (i.e. continuous) group membership, more than two groups, etc. We are particularly interested in building another layer onto the model which accounts for the "ideology" of the agents, and exploring the relationship between ideological division and polarisation based on group membership. The basic results established here will provide a base for these extensions, and we hope they will generate new interest in Schelling segregation in networks.

\bibliography{refs}

\begin{thebibliography}{}

\bibitem[Anastasi and Dalla~Riva, 2025a]{Anastasi_dimensionality_2025}
Anastasi, S. and Dalla~Riva, G. (2025a).
\newblock Measuring changes in polarisation using singular value decomposition
  of network graphs.
\newblock {\em ArXiV}.

\bibitem[Anastasi and Dalla~Riva, 2025b]{Anastasi_repository}
Anastasi, S. and Dalla~Riva, G. (2025b).
\newblock Schelling segregation dynamics in densely-connected social network
  graphs (code \& data).

\bibitem[Athreya et~al., 2017]{CareyPriebeSurvey}
Athreya, A., Fishkind, D.~E., Tang, M., Priebe, C.~E., Park, Y., Vogelstein,
  J.~T., Levin, K., Lyzinski, V., and Quin, Y. (2017).
\newblock Statistical inference on random dot product graphs: a survey.
\newblock {\em The Journal of Machine Learning Research}.

\bibitem[Bezanson et~al., 2017]{julia_article}
Bezanson, J., Edelman, A., Karpinski, S., and Shah, V.~B. (2017).
\newblock Julia: A fresh approach to numerical computing.
\newblock {\em SIAM Review}, 59(1):65--98.

\bibitem[Blair, 2023]{blair_outside_2023}
Blair, P. (2023).
\newblock Beyond racial attitudes: The role of outside options in the dynamics
  of white flight.
\newblock Working paper, National Bureau of Economic Research.

\bibitem[Clark, 1991]{clark_residential_1991}
Clark, W. (1991).
\newblock Residential preferences and neighborhood racial segregation: A test
  of the schelling segregation model.
\newblock {\em Demography}.

\bibitem[Clark and Fossett, 2008]{clark_context_2008}
Clark, W. and Fossett, M. (2008).
\newblock Understanding the social context of the schelling segregation model.
\newblock {\em Proceedings of the National Academy of Sciences}.

\bibitem[Collard and Ghetiu, 2016]{collard_landscape_2016}
Collard, P. and Ghetiu, T. (2016).
\newblock Segregation landscape: A new view on the schelling segregation space.
\newblock {\em Complex Systems}.

\bibitem[Datseris et~al., 2022]{Agents.jl}
Datseris, G., Vahdati, A.~R., and DuBois, T.~C. (2022).
\newblock Agents.jl: a performant and feature-full agent-based modeling
  software of minimal code complexity.
\newblock {\em {SIMULATION}}, 0(0).

\bibitem[Domic et~al., 2011]{domic_dynamics_2011}
Domic, N.~G., Goles, E., and Rica, S. (2011).
\newblock Dynamics and complexity of the schelling segregation model.
\newblock {\em Physical Review E—Statistical, Nonlinear, and Soft Matter
  Physics}.

\bibitem[Fagiolo et~al., 2007]{fagolio_segregation_2007}
Fagiolo, G., Valente, M., and Vriend, N.~J. (2007).
\newblock Segregation in networks.
\newblock {\em Journal of Economic Behavior \& Organization}.

\bibitem[Fossett, 2006]{fossett_preferences_2006}
Fossett, M. (2006).
\newblock Ethnic preferences, social distance dynamics, and residential
  segregation: Theoretical explorations using simulation analysis.
\newblock {\em The Journal of Mathematical Sociology}.

\bibitem[Fossett and Waren, 2005]{fossett_overlooked_2005}
Fossett, M. and Waren, W. (2005).
\newblock Overlooked implications of ethnic preferences for residential
  segregation in agent-based models.
\newblock {\em Urban Studies}.

\bibitem[Gandica et~al., 2016]{gandica_topology_2016}
Gandica, Y., Gargiulo, F., and Carletti, T. (2016).
\newblock Can topology reshape segregation patterns?
\newblock {\em Chaos, Solitons \& Fractals}.

\bibitem[Gilbert, 2002]{gilbert_varieties_2002}
Gilbert, N. (2002).
\newblock Varieties of emergence.
\newblock {\em Agent 2002 Conference: Social agents: ecology, exchange, and
  evolution, Chicago}.

\bibitem[Gretha et~al., 2018]{Gretha_network_2018}
Gretha, O.~B., Cristal, P.~M., and Mauhe, N. (2018).
\newblock Segregation in social networks: A simple schelling-like model.
\newblock In {\em 2018 IEEE/ACM International Conference on Advances in Social
  Networks Analysis and Mining (ASONAM)}.

\bibitem[Henry et~al., 2011]{Henry_network_2011}
Henry, A.~D., Prałat, P., and Zhang, C.-Q. (2011).
\newblock Emergence of segregation in evolving social networks.
\newblock {\em Proceedings of the National Academy of Sciences}.

\bibitem[Hetherington, 2009]{hetherington_review_2009}
Hetherington, M.~J. (2009).
\newblock Review article: Putting polarization in perspective.
\newblock {\em British Journal of Political Science}.

\bibitem[Ishida, 2024]{ishida_fuzzy_2024}
Ishida, A. (2024).
\newblock A fuzzy set extension of schelling's spatial segregation model.
\newblock {\em Journal of Computational Social Science}.

\bibitem[Jani, 2020]{jani_scarcity_2020}
Jani, A. (2020).
\newblock An extension of schelling's segregation model: Modeling the impact of
  individuals’ intolerance in the presence of resource scarcity.
\newblock {\em Communications in Nonlinear Science and Numerical Simulation}.

\bibitem[Laclau and Mouffe, 1985]{laclau_hegemony_1985}
Laclau, E. and Mouffe, C. (1985).
\newblock {\em Hegemony and socialist strategy}.
\newblock Verso, London.

\bibitem[Macy and Rijt, 2006]{MACY01062006}
Macy, M.~W. and Rijt, A. V.~D. (2006).
\newblock Ethnic preferences and residential segregation: Theoretical
  explorations beyond detroit.
\newblock {\em The Journal of Mathematical Sociology}.

\bibitem[Radi and Gardini, 2015]{radi_limitations_2015}
Radi, D. and Gardini, L. (2015).
\newblock Entry limitations and heterogeneous tolerances in a schelling-like
  segregation model.
\newblock {\em Chaos, Solitons \& Fractals}.

\bibitem[Schelling, 1971]{schelling_segregation_1971}
Schelling, T. (1971).
\newblock Dynamic models of segregation.
\newblock {\em The Journal of Mathematical Sociology}.

\bibitem[Vowles and Curtin, 2020]{vowles_exception_2020}
Vowles, J. and Curtin, J. (2020).
\newblock {\em A Populist Exception?: The 2017 New Zealand General Election}.
\newblock ANU Press.

\bibitem[Zhang, 2004]{zhang_dynamic_2004}
Zhang, J. (2004).
\newblock A dynamic model of residential segregation.
\newblock {\em The Journal of Mathematical Sociology}.

\bibitem[Zhu and Ghodsi, 2006]{zhu2006automatic}
Zhu, M. and Ghodsi, A. (2006).
\newblock Automatic dimensionality selection from the scree plot via the use of
  profile likelihood.
\newblock {\em Computational Statistics \& Data Analysis}, 51(2):918--930.

\end{thebibliography}

\end{document}